%% file: main.tex
\documentclass[12pt,a4paper]{article}
\usepackage[pdfstartview=FitH,colorlinks=true,linkcolor=black,anchorcolor=black,citecolor=black,urlcolor=black]{hyperref}
\usepackage[english]{babel}
\usepackage{amsmath,amssymb,titling,authblk}
\usepackage{slashed}
\usepackage{amsmath}
\usepackage{amscd}
\usepackage{mathrsfs}
\usepackage[normalem]{ulem}
\usepackage{appendix}
\usepackage{cite}

\usepackage{physics}
\usepackage{amsfonts,mathtools}

\ifx\pdfoutput\undefined
\usepackage{graphicx}
\else
\usepackage[pdftex]{graphicx}
\fi

\usepackage[font=small]{caption}
\usepackage{subcaption}
\usepackage{xcolor}

\begin{document}

\author[1]{Andrey A.~Shavrin\thanks{shavrin.andrey.cp@gmail.com}}

\affil[1]{Saint Petersburg State University, 7/9 Universitetskaya nab.\\
St. Petersburg, 199034, Russia}

\title{Holographic composite Higgs model and gravitational waves produced during first order phase transition}

\maketitle

\begin{abstract}
    The soft-wall holographic composite Higgs model assumes first-order phase transition from the dynamical inner symmetry breaking. This research focuses on the implications of the semi-analytical perturbative solution of the dual 5-dimensional theory as an effective description of the strongly coupled composite Higgs sector. We clarify the thermodynamical description and gravitational waves spectrum produced during the phase transition, which were previously numerically estimated. Besides, we investigate the limits of the applicability of our solution within the thin-wall approximation and quasiclassical approach in terms of the dual theory, that correspond to the strongly coupled regime of composite Higgs model. Our semi-analytic framework provides description of the strong first-order phase transition within the runaway scenario.
\end{abstract}

\input{Introduction}
\input{Model}
\input{Nucleation}
\input{Spectrum}
\input{Conclusion}

\section*{Acknowledgements}
We would like to express our deepest gratitude to Oleg Novikov for insightful discussions and invaluable suggestions. The work was supported by the Foundation for the Advancement of Theoretical Physics and Mathematics “BASIS” with the grant No 24-1-5-17-1.


\end{document}

%% file: Introduction.tex
\section{Introduction}


The baryon asymmetry problem, the dominance of matter over antimatter, continues to be debated since it cannot be resolved within the Standard model (SM) due to absence of first order phase transition (FOPT) according to the Sakharov conditions. They include the violation of the baryon number conservation law, the violation of C and CP symmetries and departure from thermal equilibrium. Particularly, these conditions provide motivation for models of beyond-SM physics with FOPT~\cite{Sakharov:1967dj,Sakharov:1991,White:2016nbo}.

The first two Sakharov conditions are satisfied within the SM. The first is provided by sphaleron nonperturbative processes during electroweak phase transition (PT)~\cite{Kuzmin:1985mm,Shaposhnikov:1986jp,Arnold:1987mh,Klinkhamer:1984di,Michael:2003}. The second is met due to the complex phases in the Cabibbo-Kobayashi-Maskawa and Pontecorvo-
Maki-Nakagawa-Sagata mixing matrices. However, they produce negligible contribution to the observed baryon asymmetry~\cite{khriplovich2012cp,ACME:18,chen2021heavy,zakharova2022rotating}.

The SM does not contain a mechanism for PT (only crossover). This fact justifies the need for an extension that yields a first order phase transition (FOPT). In this case, true and false vacua can exist simultaneously in different regions due to the potential barrier. The equilibrium breaking occurs at the phase boundaries leading to bubble nucleation, when the cosmic plasma interacts with the bubbles of the new phase. These processes allow production of the baryon asymmetry \cite{Zeldovich:411756,PhysRevD.15.2929}.

Part of the energy liberated during early universe FOPT converts to the gravitational waves (GW) radiation, providing a new instrument for particle physics. GW observatories place additional constraints on the physical models as well as particle colliders~\cite{kosowsky69s,turner1992wilczek,kosowsky1993gravitational,kamionkowski1994gravitational,caprini2016science,Weir:2017wfa,Geller:2018mwu,Bai:2021ibt}.


The composite Higgs (CH) model suggests an extension of the SM by replacing the\linebreak \makebox{$SU(2)_L$-Higgs} bosons with the condensate of the new strongly coupled sector with some inner symmetry $\mathcal{G}$~\cite{Contino:2003ve,Agashe:2004rs,Contino:2010rs,bellazzini2014composite,Panico:2015jxa}.
It breaks spontaneously to the subgroup $\mathcal{H}$ at a characteristic energy scale. This transition leads to the appearance of Nambu–Goldstone (NG) bosons. They become pseudo-Nambu-Goldstone (pNG) bosons after electroweak symmetry breaking. The Higgs boson is considered one of these, a composite particle of the coset $\mathcal{G}/\mathcal{H}$ analogously to the $\pi$-mesons in quantum chromodynamics (QCD) with chiral symmetry breaking. This provides the minimal realization of the symmetry with $\mathcal{G} = SO(5)\times U(1)_{B-L}$ and $\mathcal{H} = SO(4)\times U(1)_{B-L}$. The extended cosets are also studied~\cite{DaRold:2019ccj,Cheng:2020dum,Xie:2020bkl,Bian:2019kmg,Frandsen:2023vhu,Fujikura:2023fbi}, but this research focuses only on the minimal CH.

CH interacts with the fields of SM through the gauge fields related to the new inner symmetry. The terms in the total Lagrangian of CH-SM interactions can be written as
\begin{equation}
    \mathcal{L}_\text{Inter.} = B_\mu J_Y^\mu + W_\mu^k J_L^{k,\mu} + \sum_r \bar \psi_r O_r + \text{h.c.},
\end{equation}
where the SM fields: the gauge fields $B_\mu$ of $U(1)_Y$, $W_\mu^k$ of $SU(2)_L$ and fermions $\psi_r$ coupled with conserved currents $J$ and composite operators $O_r$ of CM. Here $U(1) \subset SU(2)_R \times U(1)_{B-L}$ with $SO(4) \cong SU(2)_L \times SU(2)_R$.

The spontaneous symmetry breaking is usually introduced with a chiral condensate $\Sigma_{IJ} = \langle\Psi_I\Psi_J\rangle$ of the new fundamental matter fields $\Phi_I$ with hypercolor index $I$ of the fundamental representation of $\mathcal{G}$. The generators of the coset $SO(5)/SO(4)$ labeled as $T^\alpha$ produce the NG bosons $\pi_\alpha$, while the generators of the broken symmetry $SO(4)$ denoted as $T^a$ lead to the heavy bosons, the ``radial'' fluctuations $\sigma_a$
\begin{equation}\label{eq:CHCondensate}
    \Sigma = \xi^\top \begin{pmatrix}
        0_{4\times 4} & 0\\ 0 & \Sigma_0
    \end{pmatrix}\xi + \sigma_a T^a, \quad \xi = e^{- \frac{i \pi_\alpha T^\alpha}{f_\pi}},
\end{equation}
here $f_\pi \sim \sqrt{N} \mu_\text{IR} / (4\pi)$ is the decay constant with the mass gap $\mu_{IR} \sim 1 \text{ -- }10$ TeV of the fields $\Psi_I$ and hypercolor number $N$, that is supposed to be large $N \gg 1$. 

The full symmetry $\mathcal{G}$ phase corresponds to the zeroth VEV $\Sigma_0 = 0$, while the broken symmetry $\mathcal{H}$ is defined for the nontrivial one $\Sigma_0 \neq 0$. PT between these configurations occurs at the temperature when free energy of the broken symmetry phase becomes favorable over the full symmetry one. Effective potential is mostly defined in terms of VEV within the background field approach. This research focuses only on its dynamics. The contributions of pNG or radial fluctuations are considered to be negligible.


Holographic duality, the AdS/CFT correspondence provides a non-perturbative approach to strongly coupled quantum field theories. The conjecture states the duality between the partition function of the gravitational theory with matter in the bulk of the anti de Sitter (AdS) spacetime and the generating functional of the conformal theory on the boundary. The asymptotic behavior of the bulk fields $\mathcal{X} \sim z^\Delta \mathcal{X}_0 + z^{4 - \Delta} \mathcal{X}_1$ defines the sources for the operators of the boundary theory $J \sim \mathcal{X}_0$
\begin{equation}
    \mathcal{Z}_\text{CFT}[J] = \mathcal{Z}_\text{AdS}[\mathcal{X}].
\end{equation}
Bulk fields define the symmetries, the dimensionality $\Delta$ and all other properties of the dual fields. The Lagrangian of the boundary theory is considered to be undefined in a general case (for bottom-up approach), while all the observables are studied with the correlation functions represented through the boundary term of the on-shell theory in AdS
\begin{equation}
    \langle O \ldots O \rangle = \frac{\delta^n}{\delta J^n}\mathcal{Z}_\text{CFT}
    = \left.\frac{\delta^n}{\delta \mathcal{X}_0^n}\mathcal{Z}_\text{AdS}\right|_\text{on-shell}.
\end{equation}
Practically, holography paves the way for the derivation of the VEV of the boundary fields $\langle O \rangle \sim \mathcal{X}_1$ and the free energy corresponding to the zero-point correlation function $\mathcal{F} = - T \log\mathcal{Z}$. It is the same for both theories due to holographic relation. The temperature is introduced by the AdS solutions with event horizon via the Hawking temperature~\cite{Ewerz'16}.

The main advantage of the holographic approach is the  inverse relation between coupling constants. In other words, the strongly coupled sector of the boundary theory can be treated through the weakly coupled dual theory in AdS, which can be studied in the quasiclassical approximation. It provides a powerful tool for the effective description of the quantum theory at the boundary, whereby the partition function in the holographic correspondence can be replaced with the boundary term of the on-shell action
\begin{equation}
    \left.\mathcal{Z}_\text{AdS}\right|_\text{on-shell}
    = e^{- S_{\partial\text{AdS}}}\left(1 + \mathcal{O}(\text{Vol}_3/S_{\partial\text{AdS}})\right).
\end{equation}
The boundary asymptotic behavior of the bulk classical equations of motion $S_\text{AdS}$ defines all the dynamics of the dual theory in the strong-coupling, high-energy regime.

The CH can be treated as a strongly coupled Yang–Mills theory in the large-$N$ limit. Following AdS/QCD, one can investigate the strongly coupled CH model through the dual weakly coupled theory within a bottom-up soft-wall model~\cite{Espriu:2017mlq,Katanaeva:2018frz,Espriu:2020hae,Falkowski:2008fz,Bellazzini:2014yua,Csaki:2022htl,Afonin:2022qkl,Elander:2020nyd,Elander:2021bmt,Elander:2023aow}, hard-wall approach~\cite{agashe2020cosmological,agashe2021phase}
and via top-down constructions~\cite{Erdmenger:2020lvq,Erdmenger:2020flu}.


The AdS/QCD correspondence has achieved notable success as an effective description of the strongly coupled QCD with a large number of colours. The CH model inspired by QCD can likewise be studied via a dual theory. Our model, introduced in Ref.~\cite{Novikov'22}, suggests a soft-wall bottom-up holographic description of CH. This work focuses primarily on this specific realization. However, the results may be generalized and compared to a broader class of soft-wall holographic CH models.

The holographic approach allows one to describe several distinct types of PT. The new strongly coupled sector of CH model not only influences the electroweak PT and other aspects of the dynamics of the SM fields~\cite{Chung:2021ekz,Cacciapaglia:2020jvj,Guan:2019qux,Bruggisser:2018mrt}, but also introduces some novel PTs including the confiment-deconfinment transition associated with Hawking-Page PT from the thermal AdS to the AdS black hole configuration. That is widely considered in the case of AdS/QCD~\cite{Afonin:2022qkl,Afonin'23,Chen'23,Braga'25}. In this context, holography can also effectively describe PTs in brane-world scenarios such as Randall-Sundrum~\cite{vonHarling:2017yew,Agrawal:2021alq,vonHarling:2023dfl}. The spontaneous inner symmetry breaking $\mathcal{G} \to \mathcal{H}$ with nonlinear potential should be associated with chiral PT in QCD~\cite{cherman2009chiral,gherghetta2009chiral,guralnik2011dynamics,colangelo2012temperature,li2013dynamical,he2013phase,bartz2014dynamical,chelabi2016chiral,fang2016chiral,fang2016chiral2,bartz2016chiral,fang2019chiral}. The order parameter is clearly introduced with the VEV of the CH model condensate $\Sigma_0$. It is zero in the symmetric phase, matching the trivial classical solution, and nonzero in the broken phase, corresponding to a nontrivial solution.


In a recent publication \cite{Novikov'22} we presented the model and formulated the perturbative approach for the nonlinear equations of motion of the theory in AdS. We also numerically estimated its GW spectrum. This work suggests the analytical derivation of the GW spectrum based on the perturbative solution. It also aims to clarify the limits of the applicability of our solution. It paves the way to the precise description of the thermodynamic characteristics of PT.

The following section~\ref{sec:Model} provides a review of the model and the perturbative solution. It additionally describes the temperature dependence and the applicability of the quasiclassical approach. Section~\ref{sec:Nucleation} examines key PT characteristics, including the nucleation temperature, the duration of PT via $\beta/H$ ratio and the phase transition strength $\alpha$, which represents the liberated energy. Besides, it provides a discussion of the applicability of the perturbative solution within the thin-wall approximation and the possible values of coupling constants and the masses of the radial bosons. The final section~\ref{sec:Spectrum} focuses on the GW spectrum of the considered model and possible generalizations of the solution.

%% file: Model.tex
\section{Model}\label{sec:Model}

Besides the sectors dual to SM, CH and the interaction terms, the holographic bottom-up CH model also includes the gravitational and gauge sectors. The Einstein-dilaton sector defines the gravitational dynamics and can be used for the description of the confinement/deconfinement PT via Hawking-Page transition. One can consider the dual fields to CH sector to be weakly coupled with gravity. This allows one to consider chiral PT separately with a fixed geometry of the AdS black hole with quadratic dilaton and soft-wall infrared cutoff
\begin{equation}
    ds^2 = \frac{L^2}{z^2} \left( - f(z) dt^2 + \frac{dz^2}{f(z)} + d\vec x^2\right), \quad
    f(z) = 1 - \frac{z^4}{z_\text{H}^4}, \quad \Phi = \phi_2 \frac{z^2}{z_\text{H}^2},
\end{equation}
here $L$ is the AdS radius and $z_\text{H}$ is the coordinate of the black hole horizon in the bulk. The solution defines the Hawking temperature $T = 1/(\pi z_\text{H})$.

The chiral transition occurs between different solutions of the EoM of the CH sector. At this stage we neglect the interactions with gauge fields and focus only on the transition between different solutions of the background field defined below. Gauge-mediated SM–CH interactions are suppressed, so CH dynamics decouple. The fields dual to CH can be considered separately from the others in the model.

The CH sector is described by the following Lagrangian
\begin{equation}
    \mathcal{L} = - \frac{1}{2}\tr(\partial \mathcal{X})^2 - \frac{m_\mathcal{X}^2}{2} \tr\mathcal{X}^2 + V_\mathcal{X},
\end{equation}
here the field $\mathcal{X}$ provides the sources $J$ for the VEV of the CH condensate $\Sigma$. The conformal condition on the AdS boundary requires the solution to have the asymptotic
\begin{equation}\label{eq:AsymptoticFullField}
    L \mathcal{X}_{IJ}(z,x) \sim z^\Delta \frac{\sqrt{N}}{2\pi} J_{IJ}(x) + z^{4-\Delta} \frac{\sqrt{N}}{2\pi} \Sigma_{IJ}(x) + \mathcal{O}(z^4),
\end{equation}
here $\Delta$ is the conformal dimensionality of the dual field of the boundary CH, the condensate $\Sigma$. To ensure that VEV $\Sigma_0$ has the correct dimensionality, it should be set to $\Delta = 1$. The normalization factors $\sqrt{N}/2\pi$ are given in Ref.~\cite{cherman2009chiral,Novikov'22}. The condition at the horizon is imposed by requiring the vanishing normal derivative $\partial_z \mathcal{X}|_{z = z_\text{H}} = 0$.

The mass is fixed by the conformal dimansionality of the CH field (\ref{eq:AsymptoticFullField}) through the condition derived from the asymptotic form of the EoM near the AdS boundary
\begin{equation}
    \Delta = 2 \pm \sqrt{4 + L^2m_\mathcal{X}^2}, \quad
    m_\mathcal{X}^2 = - \frac{3}{L^2}.
\end{equation}

The corresponding action in AdS with the black hole is defined by the invariant volume $\sqrt{-g} = L^5/z^5$ and integration from the boundary at $z = 0$ to the horizon at $z = z_H$
\begin{equation}
    S_\text{CH} = \frac{1}{\kappa_s} \int\limits_\text{AdS} d^4x dz \sqrt{-g} e^\Phi \mathcal{L},
\end{equation}
here $\kappa_s$ is the constant matching the mass dimensions of bulk and boundary fields. We set $\kappa_s = L$ to match the Higgs dimensionality~\cite{Katanaeva:2018frz,Espriu:2020hae}.

The quasiclassical approach provides the following simple relation between the Wick rotated Euclidean action $S_\text{E}$ and the free energy
\begin{equation}
    F = - T \log \mathcal{Z} \approx T S_\text{CH-E}.
\end{equation}
Generally, the horizon coordinate can be extracted from the action. It clarifies the temperature scaling of the free energy density
\begin{equation}
    \frac{d F}{d \text{Vol}_3} = \frac{T L^4}{z_\text{H}^4} \int_0^\beta d\tau \int_0^1 \frac{d\zeta}{\zeta^5} e^\Phi \mathcal{L}.
\end{equation}


Let us consider a specific model with a given potential as an example. The dual model should meet the full $\mathcal{G}$-symmetry at high temperatures and broken $\mathcal{H}$-symmetry in the low-temperature regime. The second requirement can be fulfilled with a nontrivial minimum (without potential barrier), that provides the vacuum solution with a non-zero VEV. For simplicity the other saddle points can be excluded with monotonous potential wall $V_\mathcal{X} \xrightarrow{\mathcal{X} \to \infty} \infty$. The simplest suitable potential that provides a first order phase transition is
\begin{equation}
    V_\mathcal{X} = - \frac{v_4}{4} \tr (\mathcal{X}^\top\mathcal{X})^2 + L^2 \frac{v_6}{6}(\mathcal{X}^\top\mathcal{X})^3,
\end{equation}
here $v_4>0$ and $v_6>0$ are dimensionless small parameters. The opposite sign leads to second order phase transition.

Within background field approach the main contribution comes from the field dual to $\Sigma_0$. One can consider the problem on the dual field neglecting the contributions of the fluctuations. In terms of the field $\mathcal{X}$ that means that we should set to zero all its components except the one that breaks the symmetry. The general approach to calculating the free energy density includes counting all the scales of the theory. For this purpose we extract the coupling constant $v_4$. It also allows one to define the sphere of applicability of the quasiclassical approach in the dual theory. The rescaling also impacts on the boundary behavior. It provides the source and VEV in terms of the rescaled constants. The original units can be recovered with the expression
\begin{equation}
    \mathcal{X} = 0_{4\times 4} \oplus \frac{\sqrt{3}}{\sqrt{v_4}L} \chi, \quad
    \chi \sim j \zeta + \sigma \zeta^3 + \mathcal{O}(\zeta^4).
\end{equation}
The asymptotic introduces the direct correspondence for the sources and the VEV
\begin{equation}
    \frac{J_0}{j} = \frac{2\pi}{z_H} \sqrt{\frac{3}{v_4 N}}, \quad
    \frac{\Sigma_0}{\sigma} = \frac{1}{2\pi z_H^3} \sqrt{\frac{3N}{v_4}}.
\end{equation}
Rescaling allows to extract all the parameters from the Lagrangian except the coupling constant ratio $\gamma = 9 v_6/v_4^2$ and the dilaton parameter $\phi_2$
\begin{equation}
    \mathcal{L} = \frac{3}{v_4 L^4} \left(- \frac{1}{2} L^2 (\partial \chi)^2 - \frac{L^2 m_\mathcal{X}^2}{2} \chi^2 - \frac{3}{4} \chi^4 + \frac{\gamma}{6}\chi^6\right),
\end{equation}
where $L^2$ in the kinetic term disappears after integration by parts.

In this research VEV $\Sigma_0$ is considered to be homogeneous. So, the dual field can depend only on the bulk coordinate $\chi(z)$. The approach simplifies EoM to
\begin{equation}\label{eq:EoM}
    \zeta^5 e^{-\Phi}\partial_\zeta \left(\frac{e^\Phi}{\zeta^3} f(z_\text{H}\zeta) \partial_\zeta \chi\right)
    - 3\chi + 3\chi^3 - \gamma\chi^5 = 0.
\end{equation}
The homogeneous solution also simplifies the free energy and defines the dimensionless free energy density $\mathcal{F}$
\begin{equation}
    F = \text{Vol}_3 \frac{6\pi}{v_4 z_\text{H}^4} \mathcal{F}, \quad
    \mathcal{F} = \int_0^1 \frac{d\zeta}{\zeta^5} \left(- \frac{1}{2} L^2 (\partial \chi)^2 - \frac{L^2 m_\mathcal{X}^2}{2} \chi^2 - \frac{3}{4} \chi^4 + \frac{\gamma}{6}\chi^6\right).
\end{equation}
According to boundary asymptotics eq.~(\ref{eq:AsymptoticFullField}), the free energy density is finite in the absence of sources $j = 0$.


EoM (\ref{eq:EoM}) can be solved with a perturbation theory with power series expansion over a small parameter $\lambda$ defined with the boundary condition at the horizon $\chi(1) = \sqrt{\lambda} \ll 1$. The solution of the nonlinear EoM (\ref{eq:EoM}) can exist in the absence  of sources $j = 0$ only for certain values of the dilaton parameter $\phi_2$ and the VEV $\sigma$ due to a specific boundary condition $\chi\sim\sigma\zeta^3$. Therefore, these parameters should be also defined via perturbative power series expansion. The zeroth order (in terms of $\lambda$) of the equation cannot be solved in the existing functions, so two linearly independent solutions were introduced as new special functions~\cite{Novikov'22}.

In this work, we consider the modified perturbation theory with the expansion, that provides a more accurate solution near the critical temperature. We introduce this approach below. This approach suggests a large coupling constant ratio $\gamma\lambda \sim 1$. We consider $\gamma > 10$ as a limit of the applicability of our perturbative approach. However, the better match with numerical calculations is reached for $\gamma \gtrsim 30$ (see Ref.~\cite{Novikov'22} for details and comparison of numerical and semi-analytical solutions).

The linear approximation \makebox{$\chi = \sqrt{\lambda} \left(\chi^{(0)} + \lambda \chi^{(1)} + \mathcal{O}(\lambda^2)\right)$} provides the perturbative free energy density, VEV and the dilaton parameter (see the derivation in \S5 and \S6 of Ref.~\cite{Novikov'22})
\begin{align}\label{eq:SemiAnalyticalSolution}
    \mathcal{F}|_{j=0} &= 2.18\lambda^2 + (-1.27 - 0.69 \gamma)\lambda^3,\\
    \phi_2 &= 2.58 - 1.68 \lambda + 0.39 \gamma\lambda^2,\\
    \sigma|_{j=0} &= \sqrt{\lambda} \left(4.41 - 3.43 \lambda + 0.95 \gamma\lambda^2\right).
\end{align}


To correctly implement the solution in the boundary theory one needs to consider the effective field theory. Its action can be defined as a Legendre transformation of the connected part of the generating functional. The free energy of the 4-dimensional theory can be considered as the connected part of the generating functional $W[J] = - \log\mathcal{Z}[J]$ \cite{Kiritsis'12}. The VEV $\Sigma_0$ defined with the field $\chi$ within the quasiclassical approach is given by
\begin{equation}
    \langle\phi\rangle = \frac{\delta W[J]}{\delta J(x)}\Big|_{J=0}.
\end{equation}
$\langle\phi\rangle$ is the classical background field. For the vacuum of the non-trivial solution $\chi\, \not\equiv\,0$, it  must be associated with the phase of broken symmetry with the nontrivial VEV $\langle\phi\rangle = \phi_- \neq 0$. The trivial solution gives the full symmetry phase $\langle\phi\rangle = \phi_+ = 0$.

The Legendre transformation provides the effective action for this VEV
\begin{equation}
    \Gamma_\text{eff} = W[J] - \int_{\text{Mink}_4} d^4x J(x) \langle\phi\rangle(x).
\end{equation}
As discussed above, we consider only the homogeneous VEV $\langle\phi\rangle = \text{const}$, where the effective action does not contain the kinetic term. It equals the effective potential with the volume a multipliers $\Gamma_\text{eff} = (\text{Vol}_3/T) V_\text{eff}$. EoM becomes the extrema conditions on the effective potential with the translation invariant potential
\begin{equation}\label{eq:EffEoM}
    \frac{\delta \Gamma_\text{eff}}{\delta \langle\phi\rangle} = J, \quad
    \frac{\partial V_\text{eff}}{\partial \langle\phi\rangle} = 0.
\end{equation}
Here we replace variation with derivation implying homogeneous VEV. The perturbative solution in the absence of sources provides the value of the free energy density at its extrema.

PT occurs between two vacua of the effective potentials, while the existence of the potential barrier provides FOPT. It makes consideration of the solutions with non-trivial source unnecessary. However, in next section we use the regularization of the action with $j \not\equiv 0$ to recover the canonical   normalization of the kinetic term, that plays the crutial role for the dynamics of the PT.


The holographic relation matches the free energies of the boundary and bulk theories. It defines the effective potential
\begin{equation}
    \left.S_\text{CH-E}\right|_{j=0}
    = \left.\Gamma_\text{eff}\right|_{J=0}.
\end{equation}
This allows one to define the effective potential in terms of the free energy density
\begin{equation}\label{eq:VEffFreeEnergyDensity}
    V_\text{eff} = \frac{6\pi}{v_4 z_\text{H}^4} \mathcal{F}.
\end{equation}


The temperature introduced as a Hawking radiation in the AdS can be related with the mass of the heavy bosons of CH associated with radial fluctuations of the condensate. The mass can be obtained as a small inhomogeneous correction to the background field
\begin{equation}\label{eq:BackgroungFluctuations}
    \chi(z) \mapsto \chi(z) + \delta \chi(z,x^\mu), \quad |\delta\chi| \ll |\chi|.
\end{equation}
Within the linear approximation by $\delta\chi$ one can write the equation of motion on the correction. The requirement for the Lorenz symmetry of the near boundary asymptotic bounds the temperature with the mass $T/m = \pi \sqrt{2/\phi_2}$ see \S8 in Ref.~\cite{Novikov'22}. The heavy bosons are the lightest predicted particles (minimal CH). That allows to impose the collider restrictions $m \gtrsim 1 \text{--} 3$ TeV \cite{DaRold'19,Carena'14}. It defines the temperature of the phase transition as $T \gtrsim 300$ GeV.

CH model includes electroweak PT, which is controlled by $\pi_a$ effective potential, that is out of scope of this research. As discussed above, the transition $\mathcal{G}\to\mathcal{H}$ occurs at higher temperatures, where sphaleron processes of SM washout the $B+L$ asymmetry. However, they do not change the $B-L$ charge. Therefore the produced asymmetry during $\mathcal{G}\to\mathcal{H}$ PT remains and can be considered as an explanation of the baryon asymmetry problem~\cite{Heeck'14}.


At the critical temperature $T_C$ the free energy of the state in the broken symmetry phase becomes lower than the one of the state with full symmetry. PT occurs for the temperatures below the critical one. Because of the potential barrier, the transition is first order. The critical temperature defined with the condition $\mathcal{F}(\phi_-,T_C) = \mathcal{F}(\phi_+,T_C)$. The last one can be set to be zero, that provides the critical temperature to be the non-trivial zero of perturbative free energy density. Its numerical value related to the mass of the radial bosons (in TeV) is given below. The potential barrier disappears at the temperature $T_{II}$, that corresponds to the trivial zero of the perturbative free energy density $\lambda = 0$ in terms of the solution~\cite{Novikov'22}
\begin{equation}
    \frac{T_C}{m \text{ TeV}} = 0.28 + \frac{0.08}{\gamma} + \mathcal{O}\left(\frac{1}{\gamma^2}\right), \quad
    \frac{T_{II}}{m \text{ TeV}} = 0.28.
\end{equation}

The nucleation temperature $T_n$, defined in the next section, is obtained from the dynamics of the bubble of broken phase and the universe expansion. This temperature must be within the range between the critical and the temperature of potential barrier disappearance. The weakness of coupling of the AdS theory leads to narrow temperature range of possible phase transition \makebox{$T_C - T_{II} = \mathcal{O}(\gamma)$}. However, we can not consider the limit. In that case the radius of the critical bubble of the broken phase would be infinite (see the discussion in sec.~\ref{sec:Nucleation}). We consider a finite big values of \makebox{$\gamma \sim 10 \text{ -- }100$}.


The quasiclassical treatment proceeds via a saddle-point expansion of the Euclidean action. In the absence of sources the action can be expressed as $S_\text{CH-E}|_{j=0} = - (\text{Vol}_3/T) V_\text{eff}[\langle\phi\rangle]$. Crucially, all quantum corrections to the solution likewise scale with the volume, so one may divide the entire loop expansion by it. It encodes the validity criteria of the quasiclassical saddle-point approximation in the smallness of loop corrections to this density. That can be generally transformed into the condition of the large effective potential $(R^3/T)|V_\text{eff}| \gg 1$. Where $R$ is the characteristic scale of the theory. In the case of the FOPT $R$ can be considered as a radius of the critical bubble, introduced in the following section. One can reformulate the criterion in the terms of the free energy density as follows
\begin{equation}\label{eq:QuasiclassValidity}
    v_4 \ll 6\pi^5 T^3 R^3 |\mathcal{F}(\phi_\pm(T),T)|.
\end{equation}

The applicability of the quasiclassical approach can be also defined with the loop corrections. However, we do not consider the formalism in this work. So, we consider the coupling constants additionally restricted with the value $v_4 < 1$ to ensure, that our solution is valid.

%% file: Nucleation.tex
\section{Nucleation}\label{sec:Nucleation}

The dynamics of bubbles of the broken-symmetry phase determine thermodynamic and hydrodynamic characteristics of the plasma. The overall evolution of the system during PT is governed by the bubble nucleation rate, the number of the bubbles produced per time and per volume units \cite{Linde'81,Guo'20}
\begin{equation}
    \Gamma = A(T) T^4 e^{-\frac{S_{3}(T)}{T}},
\end{equation}
with the dimensionless prefactor estimated as $A(T) \approx (S_3/(2\pi T))^\frac{3}{2}$~\cite{Ellis'19,Croon'23,Ivanov'22}. Here $S_{3}$ denotes the three-dimensional Euclidean on-shell action computed on the bounce solution and corresponds to the free energy cost required to form a spherical bubble of the broken-symmetry phase. The bounce solution provides the specific bubble profile that overcomes the free energy barrier, and the value evaluated on this solution yields the exponential suppression factor in the nucleation rate, thereby determining the likelihood of forming a critical bubble that can initiate the phase transition \cite{Moore'95vi}.

When the energy difference between the false and true vacua is small, the bubble wall becomes much thinner than the bubble radius. This is known as the thin-wall approximation, where the free energy of a bubble equals the three-dimensional bounce action $F(R,T) = S_3(T)$. It can be effectively decomposed into two competing contributions: a surface term and a volume term. Specifically, the free energy to form a bubble of radius $R$ is expressed as
\begin{equation}\label{eq:BubbleFreeEnergy}
    F(R,T) = 4\pi R^2 \mu - \frac{4\pi}{3} R^3 \left(V_\text{eff}(\phi_+) - V_\text{eff}(\phi_-)\right),
\end{equation}
where $\mu$ is the surface tension of the bubble wall and $V_\text{eff}(\phi_-) - V_\text{eff}(\phi_+)$ (with $V_\text{eff}(\phi_+) = 0$) is the difference in the potential energy densities between the true and false vacua. Minimizing this free energy with respect to $R$ leads to the determination of a critical radius $R_C = - 2 \mu/V_\text{eff}(\phi_-)$ and the associated free energy barrier $F_C = (16\pi/3) \, \mu^3/V_\text{eff}^2(\phi_-)$.

The surface tension can be found via the probability of the tunneling through the potential barrier. From the general EoM for the field $\phi$ one can obtain the kinetic term from the potential as $\phi' \sim \sqrt{V_\text{eff}(\phi) - V_\text{eff}(\phi_+)}$. In our model we consider the analytical solution only in the extrema of the effective potential, where the free energy density is regular in the absence of sources (\ref{eq:EffEoM}). However, the surface tension derives from the kinetic term, which calculation was avoided in the discussion of the phase structure of the model with homogeneous solution. So, the proper description requires the canonical normalization of the background field~\cite{Baratella'18,Morgante'22}. The corresponding multiplier for the kinetic term can be estimated without the explicit derivation of the renormalized solution. The nontrivial sources induce inhomogeneous corrections to the kinetic term. Their boundary behavior provides sufficient estimates for the renormalization factor. The surface tension in terms of the dimensionless density as follows
\begin{equation}\label{eq:SurfaceTension}
    \mu = \frac{3\sqrt{3}\pi^\frac{7}{2} C}{2} \frac{T^3}{v_4}
    \int\limits_0^{\langle\sigma\rangle_\text{min}} d\sigma \sqrt{2 \mathcal{F}(\sigma)}\big|_{T = T_n}.
\end{equation}
The multiplier derives from the discussed canonical renormalization and the counterterms with the numerically estimated constant $C \sim 0.1 \text{--} 1$ (we consider $C = 0.3$), the details in Ref.~\cite{Novikov'22}, and the dimentional multipliers from~(\ref{eq:VEffFreeEnergyDensity}). The integral is estimated with the factor $\langle\sigma\rangle_\text{min} \sqrt{2\mathcal{F}(\langle\sigma\rangle_\text{max})}/2$, providing sufficient accuracy for the calculations within the considered range.

The applicability of the thin-wall approximation is given by the condition according to that, the potential barrier should be larger than the potential gap between vacuua
\begin{equation}\label{eq:ThinWallApplicability}
    |\mathcal{F}(\langle\sigma\rangle_\text{min},T)| \ll \mathcal{F}(\langle\sigma\rangle_\text{max},T).
\end{equation}
Our model does not meet this condition generally. However, it fulfills the condition within a small range between the critical temperature $T_C > T > T_\mu$ and temperature $T_\mu$, at which potential barrier height equals the gap.


Nucleation temperature $T_n$ is defined with a condition that requires the rapidity of PT to be agree with the universe expansion~\cite{Megevand'16,Athron'22,Guo'20}. Within the thin-wall approximation the condition can be formulated in a folowing way~\cite{Jiang'23}
\begin{equation}\label{eq:NuclTemperatureDefinition}
    \left.\Gamma(T)/H^4(T)\right|_{T=T_n} \approx 1.
\end{equation}
Here, the Hubble constant $H(T)$ derived from the Einstein equation
\begin{equation}\label{eq:HubbleConstantDefinition}
    H^2(T) = 8\pi \rho_\text{tot}/(3m_\text{Pl}^2),
\end{equation}
with the total energy density properly defined in eq.~(\ref{eq:EnergyDensityDefinition}) and Planck mass $m_\text{Pl} = 1.2 \cdot 10^{16} \text{ TeV}$ (do not confuse with reduced $M_\text{Pl} = m_\text{Pl}/\sqrt{8\pi}$)
\cite{Ellis'19}.


\begin{figure}[h]
    \centering
    \includegraphics[width=0.9\linewidth]{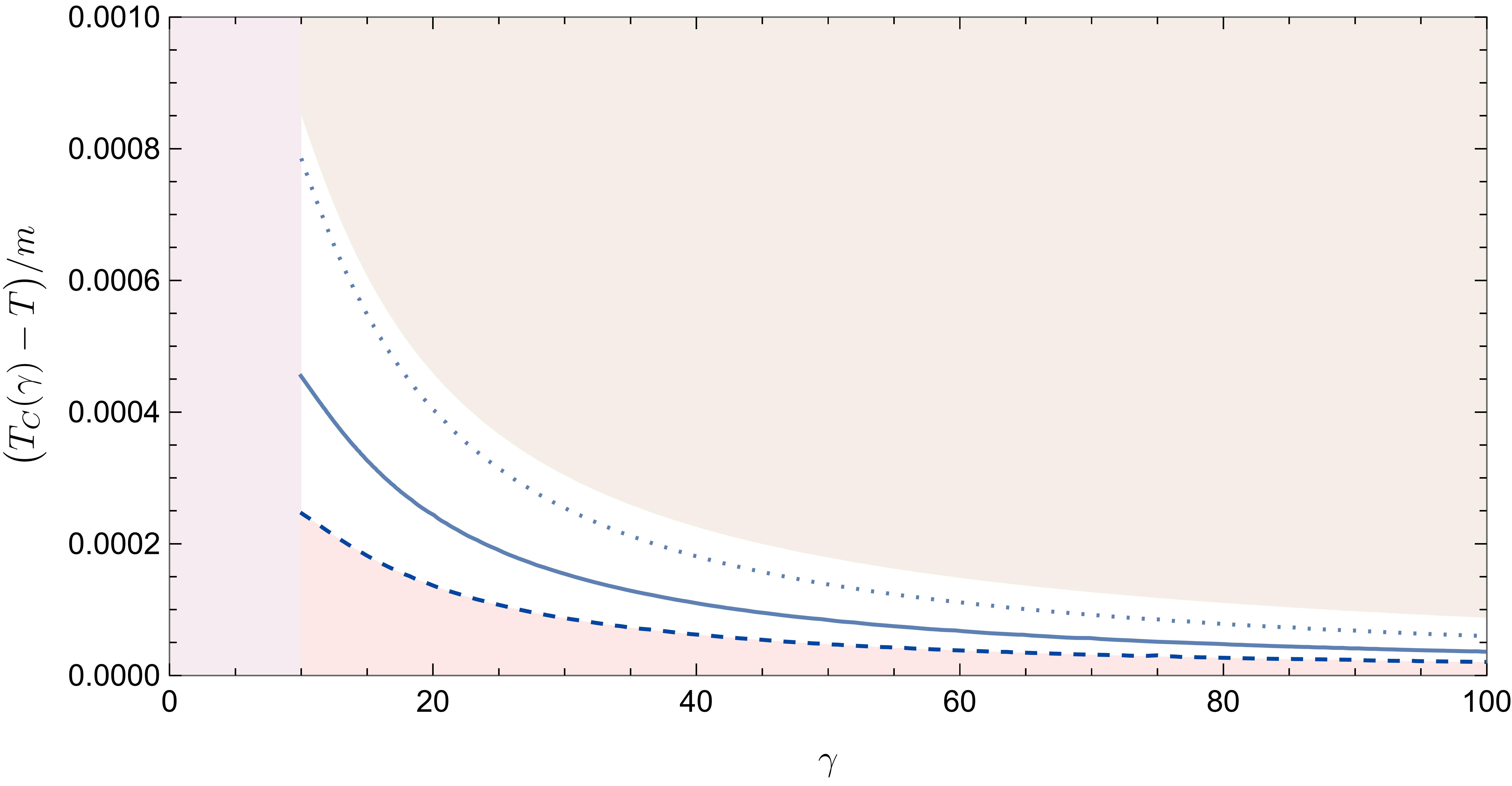}
    \caption{Nucleation temperatures and applicability of the solutions.
    Nucleation temperatures defined as $(T_C - T_n)/m$ are provided with the blue lines: dotted for $v_4 = 0.1$, solid for $v_4 = 0.3$ and dashed for $v_4 = 1$.
    The temperature range depending on $\gamma$, where thin-wall cannot be applied according to the condition eq.~(\ref{eq:ThinWallApplicability}), is denoted with a shaded light-yellow region; light-red filling denotes the region for the solutions with $v_4 >1$; light-magenta filling demonstrates the restriction of the perturbative solution $\gamma > 0$.}
    \label{fig:TNucleation}
\end{figure}

In the picture figure~\ref{fig:TNucleation} one can find the nucleation temperature defined as deviation from the critical one and related to the mass. The usage of the thin-wall approximation gives a lower boundary on the values of the coupling constant. In our case the restriction is $v_4 \gtrsim 0.1$. But we consider the calculations with this value as reliable, since the exact boundary is a bit lower. For the significantly lower values the general approach to the bubble free energy through the bounce equation should be considered within ``thick-wall'' approximation~\cite{Cutting'20,Matteini'24}. In this work we focus on the values of the coupling constant within the range where thin-wall is valid.

The quasiclassical restrictions also provide the bound on the applicability with the characteristic scale defined by the critical radius of the bubble \makebox{$R_C = - \mu/V_\text{eff}(\phi_-)$}. Since larger values of $v_4$ provide the closer temperature to the critical, the potential goes to zero \linebreak\makebox{$V_\text{eff}(\phi_-,T_n) \to V_\text{eff}(\phi_-,T_C) = 0$} at nucleation temperature despite the multiplier $1/v_4$ in eq.~(\ref{eq:VEffFreeEnergyDensity}), that redused due to the similar one in the surface tension eq.~(\ref{eq:SurfaceTension}). That makes the characteristic scale large at larger coupling constants and removes the restriction, eq.~(\ref{eq:QuasiclassValidity}), for the temperatures close to the critical one. Particularly, the quasiclassical approach is valid for the nucleation temperature within the considered range of the coupling constants. However, we keep the boundary $v_4 \ll 1$ to ensure, that the fluctuations and loop corrections are suppressed. The definition of the exact upper restriction provides the motivation for estimations of the loop corrections for the holographic solution. The unvestigations of this regime would rather be done within the traditional perturbative approach and Coleman-Weinberg potential~\cite{Liu'17,Cai'18}. Since the large coupling constant  $v_4$ of dual theory in the AdS bulk provides a weakly coupled boundary theory.


The rapidity of PT is characterized by the dimensionless $\beta/H$ ratio, where the inverse duration $\beta$ can be obtained from the evolution of the nucleation rate as $\Gamma(t) \approx \Gamma(t_n) e^{-\beta (t - t_n)}$ near the nucleation time $t_n$. $\beta$ can be defined as a linear coefficient in the Taylor expansion of the argument of the exponent
\begin{equation}
    \log \Gamma = \frac{F_C}{T} + \mathcal{O}\left(\log\frac{F_C}{T}\right),
\end{equation}
logarithmic corrections from the multiplier $A(T)$ are negligible nearly the nucleation time defined at the corresponding temperature \makebox{$F_C(T_n)/T_n|_{t=t_n} \approx 2 \log(3m_\text{Pl}^2T_n^2/(8\pi\rho_\text{tot})) \sim 140 \text{ -- } 150$} from eq.~(\ref{eq:NuclTemperatureDefinition}). It defines the inverse duration $\beta$ as follows \cite{Guo'20,Croon'23}
\begin{equation}
    \frac{F_C}{T} = \left.\frac{F_C}{T}\right|_{t = t_n} -\beta (t - t_n) + \mathcal{O}\big((t - t_n)^2\big), \quad
    \beta = - \left.\left(\frac{d}{dT}\frac{F_C}{T}\right) \frac{dT}{dt}\right|_{t = t_n}.
\end{equation}
From the evolution of the universe $dT/dt = - H T$ one can derive a more convenient formula
\begin{equation}\label{eq:BetaHRatio}
    \frac{\beta}{H} = T \left.\frac{d}{dT} \frac{F_C}{T}\right|_{T = T_n}.
\end{equation}

\begin{figure}[htbp]
    \centering
    \includegraphics[width=0.7\linewidth]{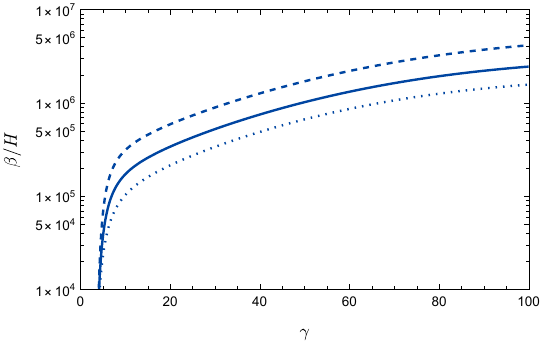}
    \caption{$\beta/H$ ratio (dotted for $v_4 = 0.1$, solid for $v_4 = 0.3$ and dashed for $v_4 = 1$).}
    \label{fig:betaH}
\end{figure}

The dependence $\beta/H \sim v_4$ in the picture figure~\ref{fig:betaH} is caused by equating the value of \makebox{$F_C/T \sim 1/v_4$} to the universe expansion eq.~(\ref{eq:NuclTemperatureDefinition}). It provides the smallest possible value for the thin-wall approach. While the quasiclassical restriction bounds the value from above. The analytical solution is considered to be valid for coupling‐constant ratio $\gamma \gtrsim 10$. So, the possible values of $\beta/H$ for this solution are in the range $10^5 \text{ -- } 5\cdot 10^6$ (with $0.1 \gtrsim v_4 \ge 1$).

As shown in section~\ref{sec:Spectrum}, larger values of the $\beta/H$ ratio correspond to suppressed GW amplitudes. That makes large $\gamma$ less attractive, since the corresponding GW spectrum is supposed to be out of sensitivity range of planned  GW observatories.


Another crucial parameter of PT is the strength, the liberated energy density related the radiation energy density. Commonly, one defines it under the perfect-fluid approximation, even though the system is out of equilibrium. The total free energy density can be defined generally. It includes the contribution of the thermal distribution of the particles in the model and the contribution of the background scalar field. The corresponding ``fluid'' and ``scalar'' components of the energy momentum tensor are provided with the equations
\begin{equation}
    T^f_{\mu\nu} = \sum_i \int \frac{d^3k}{(2\pi)^3 2E_i}2 k_\mu k_\nu f_i(k,T), \quad
    T^\phi_{\mu\nu} = \partial_\mu\langle\phi\rangle\partial_\nu\langle\phi\rangle
    - g_{\mu\nu}\left((\partial\langle\phi\rangle)^2 + V_\text{eff}(\langle\phi\rangle,T)\right),
\end{equation}
here, $f_i(t,T)$ is the thermal distributions of species $i$ particles.

The traditional way considers the contribution of the thermal distributions as a finite-temperature correction to the zero-temperature effective potential. It is perturbatively obtained as the sum of the tree potential and the Coleman-Weinberg potential \cite{Cai'18}. The holographic approach introduces the temperature as a solution with horizon in AdS. In this case the effective potential depends on the temperature.

Within perfect fluid approximation one can defined the pressure of the thermal distributions $p_f$ and the contribution of the background field
\begin{equation}
    p_f = \frac{1}{3} a(T) T^4, \quad
    p_\phi = - V_\text{eff}.
\end{equation}
Generally, $a(T)$ is a function of temperature. It encodes deviations from perfect-fluid behavior. We neglect this dependence and consider it to be a constant $a = g_* \pi^2 / 30$ with relativistic degrees of freedom $g_*(T)$~\cite{Moore'95vi}. Here and further we consider $g_* \approx 100$ as a sufficient approximation. It allows one to define the energy density through the Legendre transformation
\begin{equation}\label{eq:EnergyDensityDefinition}
    \rho = T (\partial p/\partial T) - p.
\end{equation}
It provides the corresct definition of the PT strength $\alpha = (\rho_+ - \rho_-)/\rho_+$ at nucleation temperature, or in terms of the effective potential~\cite{Schmitz'20}
\begin{equation}\label{eq:PTstrengthDefinition}
    \alpha = \left.\frac{1}{a T^4} \left(T \frac{\partial V_\text{eff}(\phi_-)}{\partial T} - V_\text{eff}(\phi_-)\right)\right|_{T = T_n}.
\end{equation}

\begin{figure}[htbp]
    \centering
    \includegraphics[width=0.7\linewidth]{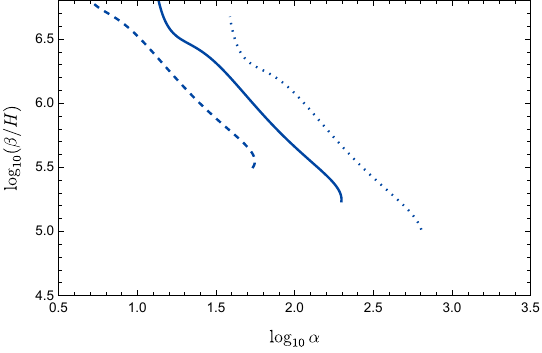}
    \caption{The dependence of the $\beta/H$ ration on PT strengh $\alpha$ (dotted for $v_4 = 0.1$, solid for $v_4 = 0.3$ and dashed for $v_4 = 1$).}
    \label{fig:alpha}
\end{figure}
One can define the dependence of $\beta/H(\alpha)$ on $\alpha$ in figure~\ref{fig:alpha}, treating the coupling ratio $\gamma$ as implicit. The larger values of $\gamma$ correspond to lower transition strength $\alpha$. Despite the wide range of $\alpha$ where the solution is valid $\alpha \sim 3 \text{ -- } 10^3$, the most relevant values are large $\alpha \gg 10$, which yield lower $\beta/H$.


The same relation can be obtained as a driving force given by the difference of the energy between vacua. It accelerates the bubble wall outward. Opposing the driving push is the frictional pressure, which arises because particles in the thermal plasma crossing the wall acquire a mass shift $\Delta m$. From a kinetic-theory perspective, one solves the Boltzmann equation for each species varying across the wall. The momentum transfer from reflected and transmitted particles yields a drag force per unit area. The friction pressure can be calculated in the relativistic limit using this mass gap~\cite{Gouttenoire'21}
\begin{equation}\label{eq:RunawayCriterion}
    \alpha_\text{fric} = \frac{1}{a T^4} \sum_i \frac{c_i g_i}{24} \Delta m_i^2 T^2.
\end{equation}
For this phase transition only the heavy bosons of the CH sector with the broken $SO(4)$ symmetry can contribute with $\Delta m \sim T_n/0.28$ and the corresponding constant $c_i = 1$. It provides the estimation $\alpha_\text{fric} \approx 0.1$, close to typical electroweak values.

When the vacuum-driven pressure exceeds all frictional forces $\alpha > \alpha_\text{fric}$ (B\"{o}deker-Moore criterion \cite{Bodeker'09}), the bubble wall never settles and its Lorentz factor grows without bound. In this runaway regime, the plasma is too feeble a drag to slow the wall, fluid disturbances are confined to an infinitesimal shock at the wall and hydrodynamic reheating ahead of the wall becomes negligible.

Conversely, if $\alpha < \alpha_\text{fric}$, friction balances the driving force and the wall reaches a terminal velocity $v_w < 1$ within a time much shorter than the phase-transition duration. In this non-runaway regime, the plasma in front of and behind the wall is strongly perturbed: a compression wave (deflagration) or rarefaction wave (detonation) forms, and substantial reheating and bulk fluid motion occur. These processes are extremely dependent on the hydrodynamics and especially on the wall speed; a brief discussion appears in section~\ref{sec:Conclusion}.

According to the dependence in figure~\ref{fig:alpha}, $\alpha \ggg \alpha_\text{fric}$, the solution corresponds to the runaway scenario with extremely ultrarelativistic wall speed and strong PT. The consideration of the non-runaway scenario $\alpha \lesssim \alpha_\text{fric}$ is complicated within considered approximations. The main restriction, complicating weak PT holographic studies, is the quasiclassical approach. It requires $v_4 < 1$ imposing a lower bound on PT strength $\alpha \propto 1/v_4$.

%% file: Spectrum.tex
\section{Spectrum}\label{sec:Spectrum}

GW can be treated as small, transverse‑traceless perturbations $h_{ij}$ on the background metric sourced by the anisotropic stress of colliding bubbles or fluid motions. These tensor perturbations satisfy a wave equation that follows from the linearized Einstein equation in the transverse-traceless gauge. It describes how the source’s quadrupole‐like stresses generate GW. Their effective energy density can be extracted as the component of the averaged energy-momentum tensor defined via the quadratic correlation function of the metric perturbation derivatives $\rho_\text{GW} = \langle\dot h_{ij}(x) \dot h_{ij}(x)\rangle/(8\pi G)$ \cite{Croon'23}.

To connect this fundamental correlator to an observable spectrum, one performs a Fourier transform of $h_{ij}$ into momentum space, defining the power spectrum of tensor perturbations $P_h(k)$ via $\langle h_{ij}(k) h_{ij}(k')\rangle \propto P_h(k)/k^3$. Replacing the time derivative with the momentum $\dot h_{ij} \sim k h_{ij}(k)$ one can substitute the power spectrum into the energy density~\cite{Santos'22}
\begin{equation}
    \rho_\text{GW} \propto \int dk k^2 (k^2 \langle h_{ij}(k) h_{ij}(k')\rangle),
\end{equation}
where the anisotropy is supposed to be negligible as well as the Universe’s expansion. The GW spectrum can be defined in a dimensionless way~\cite{Jinno'17}
\begin{equation}
    \Omega_\text{GW} = \frac{1}{\rho_c} \frac{d\rho_\text{GW}}{d \log k} \propto P_h(k),
\end{equation}
with the critical density $\rho_c(t) = 3 H^2(t) m_\text{Pl}^2/(8\pi)$ at time $t$ defined similarly to (\ref{eq:HubbleConstantDefinition}).

The spectrum can be found as~\cite{Jinno'17}
\begin{equation}\label{eq:GWSpectrum}
    \Omega_\text{GW} = \left(\frac{\beta}{H}\right)^{-2}\kappa^2 \frac{\alpha^2}{(1+\alpha)^2} S(k,\beta/H) \Delta(v_w).    
\end{equation}
The GW energy density $\rho_\text{GW}$ scales with the square of the source's energy-momentum tensor due to the quadratic nature of the Einstein field equations in perturbation theory. Therefore, it includes the square of the normalized kinetic energy fraction $\rho_\text{kin.} = \kappa \alpha \rho_\text{tot}$ with efficiency coefficient $\kappa$, which is $\kappa \approx 1$ for large PT strength $\alpha = \Delta\rho/\rho_\text{rad} \gg 1$ and $\rho_\text{tot} = (1 + \alpha) \rho_\text{rad}$. The last multiplier $(\beta/H)^{-2}$ defines the duration of the PT and can be obtained by recovering the exact relations above for bubble collisions~\cite{Kamionkowski'93}.


Spectral shape of the spectrum $S(k,\beta/H)$ in eq.~(\ref{eq:GWSpectrum}) is defined by the convolution of the source with the Green’s function of the linearized Einstein equation~\cite{Lewicki'20,Lewicki'2020a,Zhong'21}. There are three types of sources of GW during FOPT. The first one is bubble collisions, which break spherical symmetry of the bubbles, leading to nonzero quadrupole moment. Propagating bubble walls create sound waves and fluid motions in the plasma. Shock waves and magnetohydrodynamic turbulence also generate gravitational waves. However, their contributions are suppressed in the runaway regime $\alpha \ll \alpha_\text{fric}$~(\ref{eq:RunawayCriterion}), that occurs in our model. Bubble walls collide at relativistic speed without dissipation. The friction of the liberated energy transforming into the bulk fluid motion is negligible in this scenario. Therefore, bubble collisions define the GW spectrum.

Generally, the spectrum and the peak frequency depend on the wall velocity provided through \makebox{$f_*/\beta = 0.62/(v_w^2 - 0.1 v_w + 1.8)$} and $\Delta(v_w) = 0.11 v_w^3 / (0.42 + v_w^2)$ introduced in eq.~(\ref{eq:GWSpectrum}) based on simulations under the envelope approximation~\cite{Huber'08}. However, in the runaway scenario they should be considered with $v_w = 1$ due to runaway walls $\alpha \ggg \alpha_\text{fric}$.

The spectral shape is calculated numerically. For bubble collisions the fit of the spectral shape can be obtained from the behavior at high $f \gg f_0$ and low-frequency $f \ll f_0$ at large wall velocities (corresponding $\Omega_\text{GW}(f) \propto f^{-1}$ and $\Omega_\text{GW}(f) \propto f^{2.8}$)~\cite{Dev'16}
\begin{equation}
    S(f) = \frac{3.8 \left(f/f_0\right)^{2.8}}{2.8 \left(f/f_0\right)^{3.8} + 1}.
\end{equation}
Today's characteristic frequency $f_0$ is obtained by redshifting the source-frame frequency $f_* = k_*/(2\pi)$ at percolation time $t_*$ by the scale factor ratio $a_*/a_0$ with scalar factor $a(t)$ in the Friedmann-Lemaître-Robertson-Walker metric. The percolation defined by the condition that the fraction of the true vacuum of the broken phase $\phi_-$ fills approximately 30\% of the universe volume. For the regime with supercooling parameter $\delta = (T_C - T_n)/T_C \ll 1$, as one can see in figure~\ref{fig:TNucleation}, the corresponding temperatures can be considered to as equal $T_* \approx T_n$ (see figure~8 in Ref.~\cite{Athron'22}). The possible difference between nucleation and percolation times is $t_n - t_* \sim 1/\beta$ and $\sim 1/\alpha$~\cite{Megevand'16}. Therefore, based on the values of $\beta/H$ and $\alpha$ in figure~\ref{fig:alpha}, one can set $t_* \approx t_n$ in our model.

The redshift of the frequency can be found from entropy conservation $g_{s,*} T_n^3a_*^3 = g_{s,0} T_0^3 a_0^3$ with the effective number of entropy degrees of freedom $g_s \approx g$ for $T \gg 100$ GeV and $g_{s,0} \approx 4$ with today's temperature $T \approx 2.7$K~\cite{Saikawa'18,Domenech'23}. The characteristic frequency $f_*$ is given in relation to the inverse PT duration $f_*/\beta$. $\beta$ is defined within the $\beta/H_*$ ratio in eq.~(\ref{eq:BetaHRatio}). It is convenient to express the today's characteristic frequency with the expression
\begin{equation}
    f_0 = \frac{a_*}{a_0} H_0 \frac{H_*}{H_0} \frac{f_*}{\beta} \frac{\beta}{H_*},
\end{equation}
here today's Hubble constant $H_0 = 2.31 \cdot 10^{-42} h$ GeV (or $H_0 \approx 67 \, \text{km} / (\text{sec} \, \text{Mpc})$) is introduced to normalize $H_*$ with $h = 100 \, \text{sec} \, \text{Mpc} /(H_0 \, \text{km})$.

The ratio of the Hubble constants is given by eq.~(\ref{eq:HubbleConstantDefinition}) $H_*/H_0 = \sqrt{\rho_*/\rho_0}$, that usually considered with the radiation energy density $\rho_\text{rad} \propto g(T)T^4$. That substitution leads to the formula~\cite{Huber'08,Dev'16,Mohamadnejad'21,Zhong'21}
\begin{equation}
    f_0 = 1.65 \cdot 10^{-5} \frac{f_*}{\beta}
    \frac{\beta}{H_*} \frac{T_n}{0.1 \text{ TeV}} \left(\frac{g_*}{100}\right)^\frac{1}{6} \text{ Hz}
\end{equation}
where normalizations of temperature and the degrees of freedom are chosen for convenience.

The redshift factor for the spectrum is obtained from energy conservation $\rho_{\text{tot},*} a_*^4 = \rho_{\text{tot},0} a_0^4$ leading to $(a_*/a_0)^4$ from the source and the multiplier $\rho_*/\rho_0$ from the critical density. The renormalized Hubble parameter $h$ makes the spectrum $h^2\Omega$ dimensionless~\cite{Saikawa'18}. The similar calculations described above yield the spectrum of the gravitational waves with redshifted amplitude
\begin{equation}
    h^2 \Omega = 1.67 \cdot 10^{-5} \left(\frac{\beta}{H_*}\right)^{-2}
    \left(\frac{\kappa_\text{coll} \alpha}{1 + \alpha}\right)^2
    \left(\frac{g_*}{100}\right)^{-\frac{1}{3}}
    \Delta(v_w) S(f).
\end{equation}

However, when vacuum contribution $\rho_\text{vac}$ into the total energy density is not negligible (i.e. $\alpha \gtrsim 1$), the rescaling of the energies may give a more accurate result using the total density~\cite{Caprini'09}
\begin{equation}\label{eq:SpectrumFrequencyCorrection}
    \frac{\rho_{\text{tot},*}}{\rho_{\text{tot},0}}
    \approx (1 + \alpha)\frac{{\rho_{\text{rad},*}}}{\rho_{\text{rad},0}}.
\end{equation}
That leads to additional multiplier for the characteristic frequency $\sqrt{1 + \alpha}$ and spectrum $(1 + \alpha)$.

\begin{figure}[htbp]
    \centering
    \subfloat[][]{\includegraphics[width=0.49\linewidth]{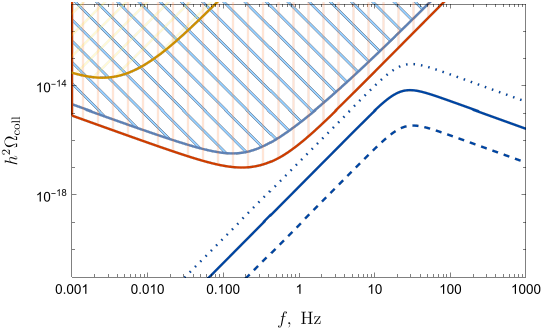}}
    ~
    \subfloat[][]{\includegraphics[width=0.49\linewidth]{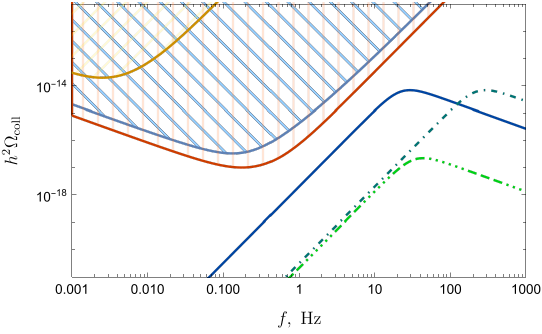}}
    
    \captionsetup{justification=raggedright,singlelinecheck=false}
    \caption{GW spectrum produced by bubble collisions $h^2\Omega(f)$: (a) blue lines-dotted for $v_4 = 0.1$ with $\alpha = 10^{2.8}$, solid for $v_4 = 0.3$ with $\alpha = 10^{2.3}$ and dashed for $v_4 = 1$ with $\alpha = 10^{1.7}$ (all at $m = 1$ TeV and $v_w = 1$); (b) dot-dashed (.-.-.-) for $v_4 = 0.3$ with $m = 1$ TeV and $v_w = 1$, `SOS' signal (...- - -...) for $v_4 = 0.3$ with $m = 10$ TeV and $v_w = 1$, ... for $v_4 = 0.3$ with $m = 1$ TeV and $v_w = 0.1$. The sensitivity curves are provided with hatched regions for the planning GW observatories: LISA, BBO, DECIGO.}
    \label{fig:Spectrums}
\end{figure}
\begin{figure}[htbp]
    \centering
    \subfloat[][]{\includegraphics[width=0.3\linewidth]{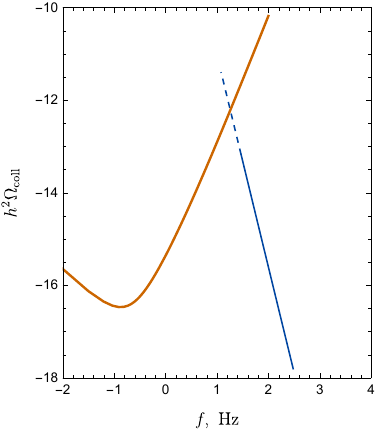}}
    ~
    \subfloat[][]{\includegraphics[width=0.3\linewidth]{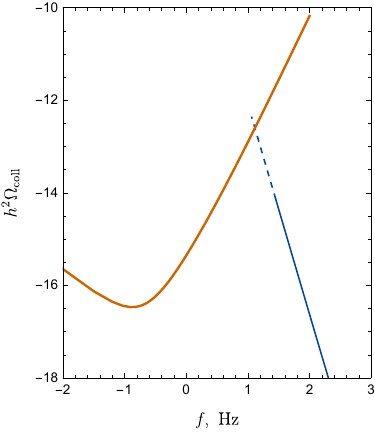}}
    ~
    \subfloat[][]{\includegraphics[width=0.3\linewidth]{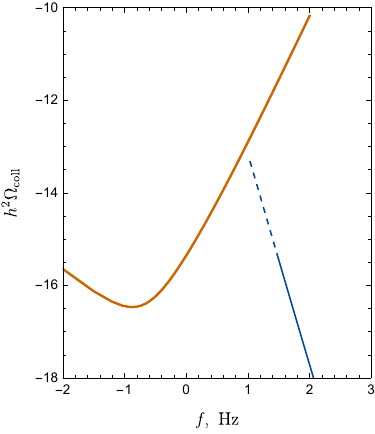}}
    
    \captionsetup{singlelinecheck=false}
    \caption{The peaks of the GW spectrum on the characteristic frequency for (a) $v_4 = 0.1$, (b) $v_4 = 0.3$, (b) $v_4 = 1$ (all at $m = 1$ TeV and $v_w = 1$); BBO sensitivity curve shown. The solid lines demonstrate the range $\alpha < \alpha_\text{max}(v_4)$. The dashed lines illustrate the exponential extrapolations $\beta/H = c_1\alpha^{c_1}$ for fitted $c_1$ and $c_2$ with the data provided with figure~\ref{fig:alpha} within the range $\alpha_\text{max}(v_4) < \alpha < 3\alpha_\text{max}(v_4)$.}
    \label{fig:Peaks}
\end{figure}

The GW spectrum is presented in figure~\ref{fig:Spectrums} (a) for different values of the coupling constant $v_4$, and figure~\ref{fig:Spectrums} (b) demonstrates the dependence of the spectrum on the mass (of the heavy bosons of CM, see the discussion after eq.~(\ref{eq:BackgroungFluctuations})) and wall speed (although our solution is not strictly valid for lower wall speeds). The peaks of the GW spectrum are presented in figure~\ref{fig:Peaks}. Solid curves show $\alpha$ within the valid range; dashed extensions extrapolate beyond the perturbative regime.

The planning GW observatories such as BBO and DECIGO almost will be able to detect the signal of the predicted predicted signal. The closest matching is provided with the large values of $\alpha \sim 10^2 \text{ -- } 10^3$, as expected from eq.~(\ref{eq:SpectrumFrequencyCorrection}). These the maximal values $\alpha_\text{max} = (10^{2.8},\, 10^{2.3},\, 10^{1.7})$ with $\alpha \sim 1/v_4$ (see eq.~(\ref{eq:VEffFreeEnergyDensity}) and eq.~(\ref{eq:PTstrengthDefinition})) correspond to the lowest value of the coupling constant ratio $\gamma \sim 10$ for $v_4 = (0.1,\, 0.3,\, 1)$. To consider a higher values one need to generalize the solution beyond the perturbative approach in eq.~(\ref{eq:SemiAnalyticalSolution}) with lower values of the coupling constant ratio $\gamma < 10$ or, alternatively, consider the whick-wall approximation with lower values of the coupling constant $v_4 < 0.1$.

%% file: Conclusion.tex
\section{Conclusion}\label{sec:Conclusion}

This publication provides the detailed semi-analytical calculations of the GW spectrum of the previously developed the holographic composite Higgs model. Particularly, the first order phase transition from the spontaneously breaking inner symmetry is considered. The strongly coupled composite Higgs is treated with the quasiclassical perturbative solution through the AdS/CFT correspondence. The research also clarifies the range of the applicability of the semi-analytical solution and provides the accurate derivations of the physical parameters previously estimated only numerically. The phase transition strength, duration and nucleation temperature are found in clear dependence of the parameters of the model and define restrictions for the applicability of the solution within the considered frameworks: quasiclassical approach of the dual theory, thin-wall approximation and perturbative approach to the equation of motion of the dual theory. Below we discuss the parameters, their method‑specific restrictions, and avenues for generalization.

Our solution provides a strong first order phase transition with large strength $\alpha \sim 3 \text{ -- } 10^3$ as shown in figure~\ref{fig:alpha}. That makes the transition extremely strong~\cite{Baldes'21,Dichtl'23} (see also Ref.~\cite{Levi'22} for weakly/strongly coupled regimes). The quasiclassical approach requires a small coupling constant $v_4 < 1$ and, therefore, limits the strength $\alpha \gtrsim 3$. It provides the framework for the strongly coupled composite Higgs model, since AdS/CFT correspondence provides the inverse proportionality between coupling constants of the bulk and boundary theories. However, the restriction $v_4 < 1$ is not explicitly defined. The criterion $S \gg 1$ formulated in eq.~(\ref{eq:QuasiclassValidity}) and discussed in section~\ref{sec:Nucleation} is generally met for our solution. The range of the quasiclassical approach validity could be clarified with estimations of the condensate fluctuations eq.~(\ref{eq:CHCondensate}). With larger values of the coupling constant $v_4$ one can consider the phase transition with lower values of the strength $\alpha$. However, too large coupling constants require loop corrections ruing the main advantage of the holographic approach through the dual theory. For this regime, the traditional perturbative QFT is more suited.

The inverse duration defined $\beta/H$ ratio happens to be large for the considered regime $\beta/H \gtrsim 10^5$ as shown in figure~\ref{fig:betaH}. It is bounded with quasiclassical and perturbative approaches eq.~(\ref{eq:SemiAnalyticalSolution}) (see \S5 and \S6 in Ref.~\cite{Novikov'22} for the details). The consideration of the lower values requires a more accurate approach to the solutions of the equation of motion of the dual theory eq.~(\ref{eq:EoM}) for lower values of $\gamma$. Thin-wall approximation also restricts $\beta/H$ ratio from below due to the dependence $\beta/H \sim v_4$ discussed in the paragraph after eq.~(\ref{eq:BetaHRatio}). The dependence can be also seen in the picture figure~\ref{fig:betaH}. The validity of the calculated nucleation temperature $v_4 \gtrsim 1$ presented in the picture figure~\ref{fig:TNucleation}. For the future investigations the thick-wall approximation can be used smaller $v_4$. However, this regime will provide even stronger transitions.

Holographic composite Higgs provides large values of the phase transition strength characteristic of primordial black hole formation models. However, their production is extremely suppressed with the large value of $\beta/H \gtrsim 10^5$ instead of $\beta/H < 7$~\cite{Gouttenoire'23,Baker'21}. Despite large $\beta/H$, the opportunity to investigate large $\alpha$ within quasiclassical approach provides motivation to apply holographic models to primordial black‑hole formation.

The non-runaway scenario for the gravitational waves does not seem to be implementable according to the criterion $\alpha < \alpha_\text{fric}$ eq.~(\ref{eq:RunawayCriterion}) for the large values of the strength. However, $\alpha_\text{fric}$ may change with the contribution of the following processes: hydrodynamic obstruction (for the thick-wall) even for $\alpha \gtrsim 1$ (see sections \S6 and \S7 in Ref.~\cite{Ai'24}) can prevent the wall speed from becoming highly relativistic~\cite{Espinosa'10}; light-to-heavy processes also give significant contributions into friction force $\propto \alpha\gamma_L^2(v_w)T^4$~\cite{Azatov'20} and $\propto \log\gamma_L(v_w)$~\cite{Ai'23a} with Lorentz factor $\gamma_L(v_w) = 1/\sqrt{1 - v_w^2} \gg 1$. At the same time, large temperature $T$ limits the friction force and provides the additional criterion $\gamma_L(v_w) \lesssim d (m/T)$ (with $d = 0.2$ for dark matter model~\cite{Jiang'23}). Therefore, the runaway criterion and $\alpha_\text{fric}$ eq.~(\ref{eq:RunawayCriterion}) for $\alpha \gg 1$ should be clarified. But at this stage we suppose runaway scenario due to extremely large $\alpha \ggg 1$ and leave the investigations of possibility of the non-runaway regime for further research.

%% file: main.bbl
\begin{thebibliography}{100}

\bibitem{Sakharov:1967dj}
A.~D. Sakharov, ``{Violation of CP Invariance, C asymmetry, and baryon asymmetry of the universe},'' {\em Pisma Zh. Eksp. Teor. Fiz.}, vol.~5, pp.~32--35, 1967.

\bibitem{Sakharov:1991}
A.~D. Sakharov, ``Violation of cp invariance, c asymmetry, and baryon asymmetry of the universe,'' {\em Soviet Physics Uspekhi}, vol.~34, pp.~392--393, may 1991.

\bibitem{White:2016nbo}
G.~A. White, ``{A Pedagogical Introduction to Electroweak Baryogenesis},'' 11 2016.

\bibitem{Kuzmin:1985mm}
V.~A. Kuzmin, V.~A. Rubakov, and M.~E. Shaposhnikov, ``{On the Anomalous Electroweak Baryon Number Nonconservation in the Early Universe},'' {\em Phys. Lett. B}, vol.~155, p.~36, 1985.

\bibitem{Shaposhnikov:1986jp}
M.~E. Shaposhnikov, ``{Possible Appearance of the Baryon Asymmetry of the Universe in an Electroweak Theory},'' {\em JETP Lett.}, vol.~44, pp.~465--468, 1986.

\bibitem{Arnold:1987mh}
P.~B. Arnold and L.~D. McLerran, ``{Sphalerons, Small Fluctuations and Baryon Number Violation in Electroweak Theory},'' {\em Phys. Rev. D}, vol.~36, p.~581, 1987.

\bibitem{Klinkhamer:1984di}
F.~R. Klinkhamer and N.~S. Manton, ``{A Saddle Point Solution in the Weinberg-Salam Theory},'' {\em Phys. Rev. D}, vol.~30, p.~2212, 1984.

\bibitem{Michael:2003}
M.~Dine and A.~Kusenko, ``Origin of the matter-antimatter asymmetry,'' {\em Rev. Mod. Phys.}, vol.~76, pp.~1--30, Dec 2004.

\bibitem{khriplovich2012cp}
I.~B. Khriplovich and S.~K. Lamoreaux, {\em CP violation without strangeness: electric dipole moments of particles, atoms, and molecules}.
\newblock Springer Science \& Business Media, 2012.

\bibitem{ACME:18}
V.~Andreev, D.~Ang, D.~DeMille, J.~Doyle, G.~Gabrielse, J.~Haefner, N.~Hutzler, Z.~Lasner, C.~Meisenhelder, B.~O'Leary, {\em et~al.}, ``Improved limit on the electric dipole moment of the electron,'' {\em Nature}, vol.~562, no.~7727, pp.~355--360, 2018.

\bibitem{chen2021heavy}
S.~Chen, Y.~Li, W.~Qian, Y.~Xie, Z.~Yang, L.~Zhang, and Y.~Zhang, ``Heavy flavour physics and cp violation at lhcb: a ten-year review,'' {\em arXiv preprint arXiv:2111.14360}, 2021.

\bibitem{zakharova2022rotating}
A.~Zakharova, ``Rotating and vibrating symmetric-top molecule raoch 3 in fundamental p, t-violation searches,'' {\em Physical Review A}, vol.~105, no.~3, p.~032811, 2022.

\bibitem{Zeldovich:411756}
Y.~B. Zel'dovich, I.~Y. Kobzarev, and L.~B. Okun, ``{Cosmological consequences of spontaneous violation of discrete symmetry},'' {\em Zh. Eksp. Teor. Fiz.}, vol.~67, pp.~3--11, 1974.

\bibitem{PhysRevD.15.2929}
S.~Coleman, ``Fate of the false vacuum: Semiclassical theory,'' {\em Phys. Rev. D}, vol.~15, pp.~2929--2936, May 1977.

\bibitem{kosowsky69s}
A.~Kosowsky, M.~S. Turner, and R.~Watkins, ``Gravitational radiation from colliding vacuum bubbles,'' {\em Phys. Rev. D}, vol.~45, pp.~4514--4535, Jun 1992.

\bibitem{turner1992wilczek}
A.~Kosowsky, M.~S. Turner, and R.~Watkins, ``Gravitational waves from first-order cosmological phase transitions,'' {\em Phys. Rev. Lett.}, vol.~69, pp.~2026--2029, Oct 1992.

\bibitem{kosowsky1993gravitational}
A.~Kosowsky and M.~S. Turner, ``Gravitational radiation from colliding vacuum bubbles: Envelope approximation to many-bubble collisions,'' {\em Physical Review D}, vol.~47, no.~10, p.~4372, 1993.

\bibitem{kamionkowski1994gravitational}
M.~Kamionkowski, A.~Kosowsky, and M.~S. Turner, ``Gravitational radiation from first-order phase transitions,'' {\em Physical Review D}, vol.~49, no.~6, p.~2837, 1994.

\bibitem{caprini2016science}
C.~Caprini, M.~Hindmarsh, S.~Huber, T.~Konstandin, J.~Kozaczuk, G.~Nardini, J.~M. No, A.~Petiteau, P.~Schwaller, G.~Servant, {\em et~al.}, ``Science with the space-based interferometer elisa. ii: Gravitational waves from cosmological phase transitions,'' {\em Journal of cosmology and astroparticle physics}, vol.~2016, no.~04, p.~001, 2016.

\bibitem{Weir:2017wfa}
D.~J. Weir, ``{Gravitational waves from a first order electroweak phase transition: a brief review},'' {\em Phil. Trans. Roy. Soc. Lond. A}, vol.~376, no.~2114, p.~20170126, 2018, 1705.01783.

\bibitem{Geller:2018mwu}
M.~Geller, A.~Hook, R.~Sundrum, and Y.~Tsai, ``{Primordial Anisotropies in the Gravitational Wave Background from Cosmological Phase Transitions},'' {\em Phys. Rev. Lett.}, vol.~121, no.~20, p.~201303, 2018, 1803.10780.

\bibitem{Bai:2021ibt}
Y.~Bai and M.~Korwar, ``{Cosmological Constraints on First-Order Phase Transitions},'' 9 2021, 2109.14765.

\bibitem{Contino:2003ve}
R.~Contino, Y.~Nomura, and A.~Pomarol, ``{Higgs as a holographic pseudoGoldstone boson},'' {\em Nucl. Phys. B}, vol.~671, pp.~148--174, 2003, hep-ph/0306259.

\bibitem{Agashe:2004rs}
K.~Agashe, R.~Contino, and A.~Pomarol, ``{The Minimal composite Higgs model},'' {\em Nucl. Phys. B}, vol.~719, pp.~165--187, 2005, hep-ph/0412089.

\bibitem{Contino:2010rs}
R.~Contino, ``{The Higgs as a Composite Nambu-Goldstone Boson},'' in {\em {Theoretical Advanced Study Institute in Elementary Particle Physics}: {Physics of the Large and the Small}}, pp.~235--306, 2011, 1005.4269.

\bibitem{bellazzini2014composite}
B.~Bellazzini, C.~Cs{\'a}ki, and J.~Serra, ``Composite higgses,'' in {\em Supersymmetry After the Higgs Discovery}, pp.~151--175, Springer, 2014.

\bibitem{Panico:2015jxa}
G.~Panico and A.~Wulzer, {\em {The Composite Nambu-Goldstone Higgs}}, vol.~913.
\newblock Springer, 2016, 1506.01961.

\bibitem{DaRold:2019ccj}
L.~Da~Rold and A.~N. Rossia, ``{The Minimal Simple Composite Higgs Model},'' {\em JHEP}, vol.~12, p.~023, 2019, 1904.02560.

\bibitem{Cheng:2020dum}
H.-C. Cheng and Y.~Chung, ``{A More Natural Composite Higgs Model},'' {\em JHEP}, vol.~10, p.~175, 2020, 2007.11780.

\bibitem{Xie:2020bkl}
K.-P. Xie, L.~Bian, and Y.~Wu, ``{Electroweak baryogenesis and gravitational waves in a composite Higgs model with high dimensional fermion representations},'' {\em JHEP}, vol.~12, p.~047, 2020, 2005.13552.

\bibitem{Bian:2019kmg}
L.~Bian, Y.~Wu, and K.-P. Xie, ``{Electroweak phase transition with composite Higgs models: calculability, gravitational waves and collider searches},'' {\em JHEP}, vol.~12, p.~028, 2019, 1909.02014.

\bibitem{Frandsen:2023vhu}
M.~T. Frandsen, M.~Heikinheimo, M.~Rosenlyst, M.~E. Thing, and K.~Tuominen, ``{Gravitational waves from SU(N)/SP(N) composite Higgs models},'' {\em JHEP}, vol.~09, p.~022, 2023, 2302.09104.

\bibitem{Fujikura:2023fbi}
K.~Fujikura, Y.~Nakai, R.~Sato, and Y.~Wang, ``{Cosmological phase transitions in composite Higgs models},'' {\em JHEP}, vol.~09, p.~053, 2023, 2306.01305.

\bibitem{Ewerz'16}
C.~Ewerz, O.~Kaczmarek, and A.~Samberg, ``{Free Energy of a Heavy Quark-Antiquark Pair in a Thermal Medium from AdS/CFT},'' {\em JHEP}, vol.~03, p.~088, 2018, 1605.07181.

\bibitem{Espriu:2017mlq}
D.~Espriu and A.~Katanaeva, ``Holographic description of $ so (5)\rightarrow so (4) $ composite higgs model,'' {\em arXiv preprint arXiv:1706.02651}, 2017.

\bibitem{Katanaeva:2018frz}
A.~Katanaeva and D.~Espriu, ``Composite higgs models: a new holographic approach,'' {\em XIII Quark Confinement and the Hadron Spectrum. 31 July-6 August 2018. Maynooth University (Confinement2018)}, p.~275, 2018.

\bibitem{Espriu:2020hae}
D.~Espriu and A.~Katanaeva, ``{Soft wall holographic model for the minimal composite Higgs boson},'' {\em Phys. Rev. D}, vol.~103, no.~5, p.~055006, 2021, 2008.06207.

\bibitem{Falkowski:2008fz}
A.~Falkowski and M.~Perez-Victoria, ``{Electroweak Breaking on a Soft Wall},'' {\em JHEP}, vol.~12, p.~107, 2008, 0806.1737.

\bibitem{Bellazzini:2014yua}
B.~Bellazzini, C.~Cs\'aki, and J.~Serra, ``{Composite Higgses},'' {\em Eur. Phys. J. C}, vol.~74, no.~5, p.~2766, 2014, 1401.2457.

\bibitem{Csaki:2022htl}
C.~Csaki, J.~Hubisz, A.~Ismail, G.~Rigo, and F.~Sgarlata, ``{a-anomalous interactions of the holographic dilaton},'' {\em Phys. Rev. D}, vol.~106, no.~5, p.~055004, 2022, 2205.15324.

\bibitem{Afonin:2022qkl}
S.~Afonin, ``{A second Higgs near 0.5 TeV from bottom-up holographic modeling of beyond the Standard Model strong sector},'' {\em Phys. Lett. B}, vol.~840, p.~137882, 2023, 2211.07500.

\bibitem{Elander:2020nyd}
D.~Elander, M.~Frigerio, M.~Knecht, and J.-L. Kneur, ``{Holographic models of composite Higgs in the Veneziano limit. Part I. Bosonic sector},'' {\em JHEP}, vol.~03, p.~182, 2021, 2011.03003.

\bibitem{Elander:2021bmt}
D.~Elander, M.~Frigerio, M.~Knecht, and J.-L. Kneur, ``{Holographic models of composite Higgs in the Veneziano limit. Part II. Fermionic sector},'' {\em JHEP}, vol.~05, p.~066, 2022, 2112.14740.

\bibitem{Elander:2023aow}
D.~Elander, A.~Fatemiabhari, and M.~Piai, ``{Towards composite Higgs: minimal coset from a regular bottom-up holographic model},'' 3 2023, 2303.00541.

\bibitem{agashe2020cosmological}
K.~Agashe, P.~Du, M.~Ekhterachian, S.~Kumar, and R.~Sundrum, ``Cosmological phase transition of spontaneous confinement,'' {\em Journal of High Energy Physics}, vol.~2020, no.~5, pp.~1--16, 2020.

\bibitem{agashe2021phase}
K.~Agashe, P.~Du, M.~Ekhterachian, S.~Kumar, and R.~Sundrum, ``Phase transitions from the fifth dimension,'' {\em Journal of High Energy Physics}, vol.~2021, no.~2, pp.~1--31, 2021.

\bibitem{Erdmenger:2020lvq}
J.~Erdmenger, N.~Evans, W.~Porod, and K.~S. Rigatos, ``{Gauge/gravity dynamics for composite Higgs models and the top mass},'' {\em Phys. Rev. Lett.}, vol.~126, no.~7, p.~071602, 2021, 2009.10737.

\bibitem{Erdmenger:2020flu}
J.~Erdmenger, N.~Evans, W.~Porod, and K.~S. Rigatos, ``{Gauge/gravity dual dynamics for the strongly coupled sector of composite Higgs models},'' {\em JHEP}, vol.~02, p.~058, 2021, 2010.10279.

\bibitem{Novikov'22}
O.~O. Novikov and A.~A. Shavrin, ``{Holographic model for the first order phase transition in the composite Higgs boson scenario},'' {\em Phys. Rev. D}, vol.~108, no.~11, p.~115011, 2023, 2209.02331.

\bibitem{Chung:2021ekz}
Y.~Chung, ``{Flavorful composite Higgs model: Connecting the B anomalies with the hierarchy problem},'' {\em Phys. Rev. D}, vol.~104, no.~11, p.~115027, 2021, 2108.08511.

\bibitem{Cacciapaglia:2020jvj}
G.~Cacciapaglia, S.~Vatani, and C.~Zhang, ``{The Techni-Pati-Salam Composite Higgs},'' {\em Phys. Rev. D}, vol.~103, p.~055001, 2021, 2005.12302.

\bibitem{Guan:2019qux}
C.-S. Guan, T.~Ma, and J.~Shu, ``{Left-right symmetric composite Higgs model},'' {\em Phys. Rev. D}, vol.~101, no.~3, p.~035032, 2020, 1911.11765.

\bibitem{Bruggisser:2018mrt}
S.~Bruggisser, B.~Von~Harling, O.~Matsedonskyi, and G.~Servant, ``{Electroweak Phase Transition and Baryogenesis in Composite Higgs Models},'' {\em JHEP}, vol.~12, p.~099, 2018, 1804.07314.

\bibitem{Afonin'23}
S.~S. Afonin, ``{Ultraviolet regularization of energy of two static sources in the bottom-up holographic approach to strong interactions},'' {\em Theor. Math. Phys.}, vol.~216, no.~3, pp.~1278--1286, 2023, 2303.03759.

\bibitem{Chen'23}
Z.-C. Chen, S.-L. Li, P.~Wu, and H.~Yu, ``{NANOGrav hints for first-order confinement-deconfinement phase transition in different QCD-matter scenarios},'' {\em Phys. Rev. D}, vol.~109, no.~4, p.~043022, 2024, 2312.01824.

\bibitem{Braga'25}
N.~R.~F. Braga and O.~C. Junqueira, ``{Holographic QCD phase diagram for a rotating plasma in the Hawking-Page approach},'' 1 2025, 2501.16446.

\bibitem{vonHarling:2017yew}
B.~von Harling and G.~Servant, ``{QCD-induced Electroweak Phase Transition},'' {\em JHEP}, vol.~01, p.~159, 2018, 1711.11554.

\bibitem{Agrawal:2021alq}
P.~Agrawal and M.~Nee, ``{Avoided deconfinement in Randall-Sundrum models},'' {\em JHEP}, vol.~10, p.~105, 2021, 2103.05646.

\bibitem{vonHarling:2023dfl}
B.~von Harling, O.~Matsedonskyi, and G.~Servant, ``{High-Temperature Electroweak Baryogenesis with Composite Higgs},'' 7 2023, 2307.14426.

\bibitem{cherman2009chiral}
A.~Cherman, T.~D. Cohen, and E.~S. Werbos, ``Chiral condensate in holographic models of qcd,'' {\em Physical Review C}, vol.~79, no.~4, p.~045203, 2009.

\bibitem{gherghetta2009chiral}
T.~Gherghetta, J.~I. Kapusta, and T.~M. Kelley, ``Chiral symmetry breaking in the soft-wall ads/qcd model,'' {\em Physical Review D}, vol.~79, no.~7, p.~076003, 2009.

\bibitem{guralnik2011dynamics}
G.~Guralnik, Z.~Guralnik, and C.~Pehlevan, ``Dynamics of the chiral phase transition from ads/cft duality,'' {\em Journal of High Energy Physics}, vol.~2011, no.~12, pp.~1--25, 2011.

\bibitem{colangelo2012temperature}
P.~Colangelo, F.~Giannuzzi, S.~Nicotri, and V.~Tangorra, ``Temperature and quark density effects on the chiral condensate: An ads/qcd study,'' {\em The European Physical Journal C}, vol.~72, no.~8, pp.~1--7, 2012.

\bibitem{li2013dynamical}
D.~Li, M.~Huang, and Q.-S. Yan, ``A dynamical soft-wall holographic qcd model for chiral symmetry breaking and linear confinement,'' {\em The European Physical Journal C}, vol.~73, no.~10, pp.~1--7, 2013.

\bibitem{he2013phase}
S.~He, S.-Y. Wu, Y.~Yang, and P.-H. Yuan, ``Phase structure in a dynamical soft-wall holographic qcd model,'' {\em Journal of High Energy Physics}, vol.~2013, no.~4, pp.~1--23, 2013.

\bibitem{bartz2014dynamical}
S.~P. Bartz and J.~I. Kapusta, ``Dynamical three-field ads/qcd model,'' {\em Physical Review D}, vol.~90, no.~7, p.~074034, 2014.

\bibitem{chelabi2016chiral}
K.~Chelabi, Z.~Fang, M.~Huang, D.~Li, and Y.-L. Wu, ``Chiral phase transition in the soft-wall model of ads/qcd,'' {\em Journal of High Energy Physics}, vol.~2016, no.~4, pp.~1--30, 2016.

\bibitem{fang2016chiral}
Z.~Fang, Y.-L. Wu, and L.~Zhang, ``Chiral phase transition and meson spectrum in improved soft-wall ads/qcd,'' {\em Physics Letters B}, vol.~762, pp.~86--95, 2016.

\bibitem{fang2016chiral2}
Z.~Fang, S.~He, and D.~Li, ``Chiral and deconfining phase transitions from holographic qcd study,'' {\em Nuclear Physics B}, vol.~907, pp.~187--207, 2016.

\bibitem{bartz2016chiral}
S.~P. Bartz and T.~Jacobson, ``Chiral phase transition and meson melting in a soft-wall ads/qcd model,'' {\em Physical Review D}, vol.~94, no.~7, p.~075022, 2016.

\bibitem{fang2019chiral}
Z.~Fang, Y.-L. Wu, and L.~Zhang, ``Chiral phase transition and qcd phase diagram from ads/qcd,'' {\em Physical Review D}, vol.~99, no.~3, p.~034028, 2019.

\bibitem{Kiritsis'12}
E.~Kiritsis and V.~Niarchos, ``{The holographic quantum effective potential at finite temperature and density},'' {\em JHEP}, vol.~08, p.~164, 2012, 1205.6205.

\bibitem{DaRold'19}
L.~Da~Rold and A.~N. Rossia, ``{The Minimal Simple Composite Higgs Model},'' {\em JHEP}, vol.~12, p.~023, 2019, 1904.02560.

\bibitem{Carena'14}
M.~Carena, L.~Da~Rold, and E.~Pont\'on, ``{Minimal Composite Higgs Models at the LHC},'' {\em JHEP}, vol.~06, p.~159, 2014, 1402.2987.

\bibitem{Heeck'14}
J.~Heeck, ``{Unbroken B \textendash{} L symmetry},'' {\em Phys. Lett. B}, vol.~739, pp.~256--262, 2014, 1408.6845.

\bibitem{Linde'81}
A.~D. Linde, ``{Decay of the False Vacuum at Finite Temperature},'' {\em Nucl. Phys. B}, vol.~216, p.~421, 1983.
\newblock [Erratum: Nucl.Phys.B 223, 544 (1983)].

\bibitem{Guo'20}
H.-K. Guo, K.~Sinha, D.~Vagie, and G.~White, ``{Phase Transitions in an Expanding Universe: Stochastic Gravitational Waves in Standard and Non-Standard Histories},'' {\em JCAP}, vol.~01, p.~001, 2021, 2007.08537.

\bibitem{Ellis'19}
J.~Ellis, M.~Lewicki, J.~M. No, and V.~Vaskonen, ``{Gravitational wave energy budget in strongly supercooled phase transitions},'' {\em JCAP}, vol.~06, p.~024, 2019, 1903.09642.

\bibitem{Croon'23}
D.~Croon, ``{TASI lectures on Phase Transitions, Baryogenesis, and Gravitational Waves},'' {\em PoS}, vol.~TASI2022, p.~003, 2024, 2307.00068.

\bibitem{Ivanov'22}
A.~Ivanov, M.~Matteini, M.~Nemev\v{s}ek, and L.~Ubaldi, ``{Analytic thin wall false vacuum decay rate},'' {\em JHEP}, vol.~03, p.~209, 2022, 2202.04498.
\newblock [Erratum: JHEP 07, 085 (2022), Erratum: JHEP 11, 157 (2022)].

\bibitem{Moore'95vi}
G.~D. Moore and T.~Prokopec, ``{How fast can the wall move? A Study of the electroweak phase transition dynamics},'' {\em Phys. Rev. D}, vol.~52, pp.~7182--7204, 1995, hep-ph/9506475.

\bibitem{Baratella'18}
P.~Baratella, A.~Pomarol, and F.~Rompineve, ``{The Supercooled Universe},'' {\em JHEP}, vol.~03, p.~100, 2019, 1812.06996.

\bibitem{Morgante'22}
E.~Morgante, N.~Ramberg, and P.~Schwaller, ``{Gravitational waves from dark SU(3) Yang-Mills theory},'' {\em Phys. Rev. D}, vol.~107, no.~3, p.~036010, 2023, 2210.11821.

\bibitem{Megevand'16}
A.~Megevand and S.~Ramirez, ``{Bubble nucleation and growth in very strong cosmological phase transitions},'' {\em Nucl. Phys. B}, vol.~919, pp.~74--109, 2017, 1611.05853.

\bibitem{Athron'22}
P.~Athron, C.~Bal\'azs, and L.~Morris, ``{Supercool subtleties of cosmological phase transitions},'' {\em JCAP}, vol.~03, p.~006, 2023, 2212.07559.

\bibitem{Jiang'23}
S.~Jiang, A.~Yang, J.~Ma, and F.~P. Huang, ``{Implication of nano-Hertz stochastic gravitational wave on dynamical dark matter through a dark first-order phase transition},'' {\em Class. Quant. Grav.}, vol.~41, no.~6, p.~065009, 2024, 2306.17827.

\bibitem{Cutting'20}
D.~Cutting, E.~G. Escartin, M.~Hindmarsh, and D.~J. Weir, ``{Gravitational waves from vacuum first order phase transitions II: from thin to thick walls},'' {\em Phys. Rev. D}, vol.~103, no.~2, p.~023531, 2021, 2005.13537.

\bibitem{Matteini'24}
M.~Matteini, M.~Nemev\v{s}ek, Y.~Shoji, and L.~Ubaldi, ``{False vacuum decay rate from thin to thick walls},'' {\em JHEP}, vol.~04, p.~120, 2025, 2404.17632.

\bibitem{Liu'17}
D.~Liu, I.~Low, and C.~E.~M. Wagner, ``{Modification of Higgs Couplings in Minimal Composite Models},'' {\em Phys. Rev. D}, vol.~96, no.~3, p.~035013, 2017, 1703.07791.

\bibitem{Cai'18}
R.-G. Cai and S.-J. Wang, ``{Energy budget of cosmological first-order phase transition in FLRW background},'' {\em Sci. China Phys. Mech. Astron.}, vol.~61, p.~080411, 2018, 1803.03002.

\bibitem{Schmitz'20}
K.~Schmitz, ``{New Sensitivity Curves for Gravitational-Wave Signals from Cosmological Phase Transitions},'' {\em JHEP}, vol.~01, p.~097, 2021, 2002.04615.

\bibitem{Gouttenoire'21}
Y.~Gouttenoire, R.~Jinno, and F.~Sala, ``{Friction pressure on relativistic bubble walls},'' {\em JHEP}, vol.~05, p.~004, 2022, 2112.07686.

\bibitem{Bodeker'09}
D.~Bodeker and G.~D. Moore, ``{Can electroweak bubble walls run away?},'' {\em JCAP}, vol.~05, p.~009, 2009, 0903.4099.

\bibitem{Santos'22}
R.~R. L.~d. Santos and L.~M. van Manen, ``{Gravitational waves from the early universe},'' 12 2022, 2212.05594.

\bibitem{Jinno'17}
R.~Jinno, S.~Lee, H.~Seong, and M.~Takimoto, ``{Gravitational waves from first-order phase transitions: Towards model separation by bubble nucleation rate},'' {\em JCAP}, vol.~11, p.~050, 2017, 1708.01253.

\bibitem{Kamionkowski'93}
M.~Kamionkowski, A.~Kosowsky, and M.~S. Turner, ``{Gravitational radiation from first order phase transitions},'' {\em Phys. Rev. D}, vol.~49, pp.~2837--2851, 1994, astro-ph/9310044.

\bibitem{Lewicki'20}
M.~Lewicki and V.~Vaskonen, ``{Gravitational wave spectra from strongly supercooled phase transitions},'' {\em Eur. Phys. J. C}, vol.~80, no.~11, p.~1003, 2020, 2007.04967.

\bibitem{Lewicki'2020a}
M.~Lewicki and V.~Vaskonen, ``{Gravitational waves from colliding vacuum bubbles in gauge theories},'' {\em Eur. Phys. J. C}, vol.~81, no.~5, p.~437, 2021, 2012.07826.
\newblock [Erratum: Eur.Phys.J.C 81, 1077 (2021)].

\bibitem{Zhong'21}
H.~Zhong, B.~Gong, and T.~Qiu, ``{Gravitational waves from bubble collisions in FLRW spacetime},'' {\em JHEP}, vol.~02, p.~077, 2022, 2107.01845.

\bibitem{Huber'08}
S.~J. Huber and T.~Konstandin, ``{Gravitational Wave Production by Collisions: More Bubbles},'' {\em JCAP}, vol.~09, p.~022, 2008, 0806.1828.

\bibitem{Dev'16}
P.~S.~B. Dev and A.~Mazumdar, ``{Probing the Scale of New Physics by Advanced LIGO/VIRGO},'' {\em Phys. Rev. D}, vol.~93, no.~10, p.~104001, 2016, 1602.04203.

\bibitem{Saikawa'18}
K.~Saikawa and S.~Shirai, ``{Primordial gravitational waves, precisely: The role of thermodynamics in the Standard Model},'' {\em JCAP}, vol.~05, p.~035, 2018, 1803.01038.

\bibitem{Domenech'23}
G.~Dom\`enech, ``{Lectures on Gravitational Wave Signatures of Primordial Black Holes},'' 7 2023, 2307.06964.

\bibitem{Mohamadnejad'21}
A.~Mohamadnejad, ``{Electroweak phase transition and gravitational waves in a two-component dark matter model},'' {\em JHEP}, vol.~03, p.~188, 2022, 2111.04342.

\bibitem{Caprini'09}
C.~Caprini, R.~Durrer, T.~Konstandin, and G.~Servant, ``{General Properties of the Gravitational Wave Spectrum from Phase Transitions},'' {\em Phys. Rev. D}, vol.~79, p.~083519, 2009, 0901.1661.

\bibitem{Baldes'21}
I.~Baldes, S.~Blasi, A.~Mariotti, A.~Sevrin, and K.~Turbang, ``{Baryogenesis via relativistic bubble expansion},'' {\em Phys. Rev. D}, vol.~104, no.~11, p.~115029, 2021, 2106.15602.

\bibitem{Dichtl'23}
M.~Dichtl, J.~Nava, S.~Pascoli, and F.~Sala, ``{Baryogenesis and leptogenesis from supercooled confinement},'' {\em JHEP}, vol.~02, p.~059, 2024, 2312.09282.

\bibitem{Levi'22}
N.~Levi, T.~Opferkuch, and D.~Redigolo, ``{The supercooling window at weak and strong coupling},'' {\em JHEP}, vol.~02, p.~125, 2023, 2212.08085.

\bibitem{Gouttenoire'23}
Y.~Gouttenoire and T.~Volansky, ``{Primordial black holes from supercooled phase transitions},'' {\em Phys. Rev. D}, vol.~110, no.~4, p.~043514, 2024, 2305.04942.

\bibitem{Baker'21}
M.~J. Baker, M.~Breitbach, J.~Kopp, and L.~Mittnacht, ``{Detailed calculation of primordial black hole formation during first-order cosmological phase transitions},'' {\em Phys. Rev. D}, vol.~111, no.~6, p.~063544, 2025, 2110.00005.

\bibitem{Ai'24}
W.-Y. Ai, X.~Nagels, and M.~Vanvlasselaer, ``{Criterion for ultra-fast bubble walls: the impact of hydrodynamic obstruction},'' {\em JCAP}, vol.~03, p.~037, 2024, 2401.05911.

\bibitem{Espinosa'10}
J.~R. Espinosa, T.~Konstandin, J.~M. No, and G.~Servant, ``{Energy Budget of Cosmological First-order Phase Transitions},'' {\em JCAP}, vol.~06, p.~028, 2010, 1004.4187.

\bibitem{Azatov'20}
A.~Azatov and M.~Vanvlasselaer, ``{Bubble wall velocity: heavy physics effects},'' {\em JCAP}, vol.~01, p.~058, 2021, 2010.02590.

\bibitem{Ai'23a}
W.-Y. Ai, ``{Logarithmically divergent friction on ultrarelativistic bubble walls},'' {\em JCAP}, vol.~10, p.~052, 2023, 2308.10679.

\end{thebibliography}
